\documentclass[journal=jcisd8,manuscript=article,layout=traditional]{achemso}
\setkeys{acs}{articletitle = true}
\setkeys{acs}{etalmode = truncate}
\setkeys{acs}{maxauthors = 0}

\usepackage[version=3]{mhchem} 

\usepackage{xr-hyper}
\makeatletter
\newcommand*{\addFileDependency}[1]{
\typeout{(#1)}
\@addtofilelist{#1}
\IfFileExists{#1}{}{\typeout{No file #1.}}
}\makeatother

\usepackage{siunitx, amsmath, bbold, dsfont, hyperref, multirow, subcaption, graphicx}

\setkeys{acs}{doi = true}
\usepackage{xcolor}
\hypersetup{
    colorlinks,
    linkcolor={teal},
    citecolor={teal},
    urlcolor={teal}
}
\SectionNumbersOn

\usepackage{algorithm, fancyvrb, listings}
\usepackage{cuted}

\newcommand{\textsub}[1]{_{\text{#1}}}
\newcommand{\textsup}[1]{^{\text{#1}}}
\newcommand{\doi}[1]{\href{http://dx.doi.org/#1}{\nolinkurl{#1}}}
\DeclareSIUnit\angstrom{\text {Å}}
 \DeclareSIUnit\hartree{\text {\ensuremath {E}}_{\mathrm {h}}}
 
\author{Andrea Levy}
\affiliation{Laboratory of Computational Chemistry and Biochemistry, Ecole Polytechnique Fédérale de Lausanne, Lausanne, Switzerland}
\author{Andrej Antal\'{i}k}
\affiliation{Laboratory of Computational Chemistry and Biochemistry, Ecole Polytechnique Fédérale de Lausanne, Lausanne, Switzerland}
\author{J\'{o}gvan Magnus Haugaard Olsen}
\affiliation{DTU Chemistry, Technical University of Denmark, Lyngby, Denmark}
\author{Ursula Rothlisberger}
\affiliation{Laboratory of Computational Chemistry and Biochemistry, Ecole Polytechnique Fédérale de Lausanne, Lausanne, Switzerland}
\email{ursula.roethlisberger@epfl.ch}

\title{OpenMM--MiMiC Interface for Efficient and Flexible Multiscale Simulations}

\makeatletter
\if@twocolumn
\let\oldmaketitle\maketitle
\let\maketitle\relax
\fi
\makeatother

\begin{document}







 
\newcommand*{\abstracttext}{%
MiMiC is a flexible and efficient framework for multiscale simulations in which different subsystems are treated by individual client programs.
In this work, we present a new interface with OpenMM to be used as an MM client program and we demonstrate its efficiency for QM/MM MD simulations.
Apart from its high performance, especially on GPUs, and a wide selection of features, OpenMM is a highly-flexible and easily-extensible program, ideal for the development of novel multiscale methods.
Thanks to the open-ended design of MiMiC, the OpenMM--MiMiC interface will automatically support any new QM client program interfaced with MiMiC for QM/MM and, with minimal changes needed, new multiscale methods implemented, opening up new research directions beyond electrostatic embedding QM/MM. 
}

\makeatletter
\if@twocolumn
    \twocolumn[
    \begin{@twocolumnfalse}
    \oldmaketitle
    \begin{abstract}
        \abstracttext
    \end{abstract}
    \end{@twocolumnfalse}
    ]
\else
    \begin{abstract}
        \abstracttext
    \end{abstract} 
\fi
\makeatother

\section{Introduction}

Multiscale techniques represent an indispensable part of computational chemistry, as they provide important insights into phenomena spanning extended space and time scales.
These methods operate on the simple premise that the system can be partitioned into subsystems, each described at a different resolution or level of theory, with the additional requirement to account for the interactions between them.
The most prominent example is the hybrid quantum mechanics/molecular mechanics (QM/MM) approach\cite{warshel1976theoretical, Singh_QMMM, Karplus_QMMM, Senn_QMMM, Rothlisberger_review, Estrin_review, Mennucci_review}, in which a system is divided into one subsystem treated at the quantum mechanical (QM) level, and another treated with a classical molecular mechanics (MM) approach involving an empirical force field (FF).
Hybrid QM/MM approaches enable the description of a subset of atoms at QM resolution, without a need to apply these computationally demanding methods to the entire system.
Among the typical applications are, e.g., enzymes in which a reaction occurs at the active site, which inherently requires QM treatment to capture the process, while accounting for the effects of the environment at a lower resolution.

The software infrastructure for multiscale simulations can be implemented in a multitude of different ways.
For example, many quantum chemistry packages include QM/MM capabilities, either via direct inclusion of MM methods or interfaces to MM programs.
A different approach is to write a modular integrative framework that interfaces several MM and QM programs, handles communication, and computes subsystem interactions enabling high flexibility in the coupling of diverse programs.
However, the performance of such loosely-coupled approaches often suffers due to overheads stemming from inefficient data exchange and repeated start-ups and shutdowns of external programs.
MiMiC\cite{olsen2019mimic, viacheslav2019extreme, antalik2024mimic} is a recent framework for multiscale simulations that combines the best of both worlds. 
It has the flexibility of using a variety of QM and MM programs, yet implemented in a way that does not substantially impede efficiency.
During a multiscale simulation, individual subsystems are assigned to different external programs, while MiMiC facilitates communication among them and, at the same time, calculates the subsystems interactions.
To achieve smooth communication between a diverse range of programs, it adapts a client--server communication strategy combined with a multiple-program multiple-data (MPMD) model.
The external programs are loosely coupled to MiMiC via a simple well-defined API of its communication library, MCL.
This makes it easy to introduce new features through the introduction of new client programs, with minimal changes required on their side.

The current release of MiMiC supports QM/MM molecular dynamics (MD) simulations, within an electrostatic embedding scheme, with the QM subsystem being handled by CPMD\cite{cpmd_free} while the MM subsystem is described by classical FFs using GROMACS\cite{abraham2015,gromacs2015,pall2020heterogeneous}.
Here, we present an extension of MiMiC's MM capabilities by introducing a new client program for this purpose, namely, OpenMM\cite{eastman2013openmm,eastman2017openmm,eastman2023openmm}.
It is an open-source MD package that focuses on extensibility and efficiency across multiple platforms, particularly those based on graphics processing units (GPUs).
OpenMM supports a broad range of popular FFs, including, e.g., AMBER\cite{maier2015ff14sb} and CHARMM\cite{best2012optimization}, with the additional possibility of specifying arbitrary analytic forms for custom forces.
It also supports implicit solvent models\cite{hawkins1995pairwise, onufriev2004exploring, mongan2007generalized, nguyen2013improved}, polarizable FFs, such as AMOEBA\cite{Ponder2010-nr} and CHARMM polarizable FF\cite{lopes2013polarizable}, and more recently, also machine learning (ML) potentials have been introduced \cite{eastman2023openmm}.

Over the years, several QM/MM implementations using OpenMM have been developed.
A notable example is PyDFT-QMMM\cite{Pederson2024}, a recently published package for performing QM/MM MD simulations that interfaces OpenMM with Psi4\cite{turney2012psi4,parrish2017psi4} via a Python API.
The same two programs are also interfaced in JANUS\cite{zhang2019janus}, a package that supports adaptive QM/MM. 
PyMM\cite{chen2022pymm} takes a different approach by implementing the perturbed matrix QM/MM method, an a posteriori correction to snapshots from classical MD trajectories.
The MolSSI Driver Interface\cite{barnes2024molssi, barnes2021molssi, github_molssi} implements a subtractive QM/MM scheme. 
Finally, several projects are currently still in the development phase and are not yet officially released, among them PSIOMM\cite{github_psiomm} (OpenMM--Psi4 interface), 
an interface to OpenMM implemented in Q-Chem \cite{github_qchem},
and ASH\cite{github_ash}, a multiscale modeling software package interfacing multiple QM programs and OpenMM for QM/MM MD simulations.

In this article, we interface OpenMM with the MiMiC framework, offering several advantages to both projects.
OpenMM is an ideal client program for the development of novel multiscale methods, given its flexibility and extensibility.
This interface unlocks a new MM client program in MiMiC, making it even more flexible by providing access to numerous OpenMM features, be it those that are currently supported or the ones introduced in the future.
Conversely, new developments in MiMiC will be instantly accessible by OpenMM with no or minimal changes needed to the interface presented here.
These developments include those directly implemented in MiMiC, those accessible via an interface to a new external program, or even those newly implemented in an external program that has already been interfaced with MiMiC.
Moreover, this new interface represents a fundamental first step towards more sophisticated multiscale models such as fully-polarizable embedding QM/MM, which is currently under development in MiMiC\cite{sonata_polqmmm, antalik2024mimic}.

In the following, we illustrate the main characteristics of the MiMiC framework and OpenMM (Sec.~\ref{sec:methods}), give details of the implementation of this new interface (Sec.~\ref{sec:implementation}), and then examine its accuracy and performance for different test cases (Sec.~\ref{sec:systems} and \ref{sec:results}).

\section{\label{sec:methods}Methods}

\subsection{\label{sec:mimic_framework}MiMiC Framework}
To enable multiscale simulations with multiple external programs, MiMiC\cite{olsen2019mimic, viacheslav2019extreme, antalik2024mimic} performs two main tasks.
It facilitates the communication between the programs that treat individual subsystems and computes the subsystem interactions.
For this purpose and the sake of simplicity, it is distributed with a two-sided API.
The main MiMiC library links to an MD driver program, which is responsible for the integration of the equations of motion of the entire system, and provides the infrastructure for computing subsystem interactions.
The communication is handled with a communication library, MCL, that interfaces with external programs computing the subsystem contributions and, through its simple API, enables efficient network-based communication with minimal impact on the parallelization of these programs.
MCL's API consists of a limited number of procedures that are executed along the simulation.
The different calls are coordinated using a request-based approach: after initialization, an external client enters the main loop, referred to as the \emph{MiMiC loop}, which performs two actions for each iteration.
First, a request is received via \verb|MCL_Recv()|, then the action specified by the request is executed.

A detailed overview of the MiMiC framework, with an extensive discussion of the software design, can be found in Ref.~\citenum{antalik2024mimic}.
MiMiC is a free and open-source software and its source code is hosted on GitLab\cite{mimic-projects}.

\subsection{\label{sec:qmmm_mimic}QM/MM electrostatic embedding in MiMiC}
At the moment, MiMiC supports an additive electrostatic-embedding QM/MM scheme\cite{olsen2019mimic, antalik2024mimic}, in which the point charges of the MM atoms polarize the QM region, but with no back-polarization of the MM region by the QM one.
In terms of dynamics, both the MM atoms and QM nuclei are treated as point particles that are propagated according to the classical equations of motion, with the latter being treated via  Born--Oppenheimer dynamics, in which the (time-independent) electronic problem is solved for fixed nuclear positions at each time step, including the effect of the surrounding MM subsystem as an external potential.

The QM/MM interaction contains different terms that are computed either by the MM client program or by MiMiC.
Bonded terms, such as those describing bond stretching, bending, and torsions that include MM atoms and at least one QM atom, are treated classically and are thus computed by the MM client program.
Similarly, the van der Waals (vdW) component of the QM/MM non-bonded contributions is treated with MM. 
In contrast, the Coulomb interactions are computed by MiMiC, which currently offers a generalized version of the efficient electrostatic coupling scheme by \citeauthor{laio2002hamiltonian}\cite{laio2002hamiltonian}.
Here, Coulomb interactions are split into short-range (sr) and long-range (lr) contributions with only the former computed exactly, as
\makeatletter
\if@twocolumn
    \begin{equation}
    \label{eq:sr}
    \begin{split}
        V\textsub{QM/MM}^{sr} &= \sum_{i=1}^{N\textsub{MM}\textsup{sr}} q_i 
    \left( \int \mathrm{d} \mathbf{r}\,   
        \rho(\mathbf{r}) v\textsup{mod}(\mathbf{r}, \mathbf{R}\textsub{MM,\,$i$})\right.\\
        &\quad+\sum_{j=1}^{N\textsub{QM}} \left.\vphantom{\int} Z_{j} v\textsup{mod}(\mathbf{R}\textsub{QM,\,$j$}, \mathbf{R}\textsub{MM,\,$i$})   
        \right) \, ,
    \end{split}
    \end{equation}
\else
    \begin{equation}
    \label{eq:sr}
     V\textsub{QM/MM}\textsup{sr} = \sum_{i=1}^{N\textsub{MM}\textsup{sr}} q_i 
    \left( \int \mathrm{d} \mathbf{r}\,   
        \rho(\mathbf{r}) v\textsup{mod}(\mathbf{r}, \mathbf{R}\textsub{MM,\,$i$})
        +\sum_{j=1}^{N\textsub{QM}} \vphantom{\int} Z_{j} v\textsup{mod}(\mathbf{R}\textsub{QM,\,$j$}, \mathbf{R}\textsub{MM,\,$i$})   
        \right) \, ,
    \end{equation}
\fi
\makeatother
where $q_i$ is the point charge of the $i$th MM atom, $\rho$ is the electron density, and $Z_j$ is the charge of the $j$th QM nucleus.
To prevent electron spill-out\cite{laio2002hamiltonian}, the electrostatic interaction potential between two points $a$ and $b$ includes a function that modifies the Coulomb potential at short-range
\begin{equation}
     v\textsup{mod}(\mathbf{R}\textsub{$a$}, \mathbf{R}\textsub{$b$}) =
     \frac{r\textsub{cov,$a$}^4 - \left|\mathbf{R}\textsub{$b$} - \mathbf{R}\textsub{$a$}\right|^4}
     {r\textsub{cov,$a$}^5 - \left|\mathbf{R}\textsub{$b$} - \mathbf{R}\textsub{$a$}\right|^5} \, ,
\end{equation}
where $r\textsub{cov,$a$}$ is the covalent radius of atom $a$.

To account for long-range interactions, the MM point charges interact with a multipole expansion of the QM electrostatic potential around the origin $\bar{\mathbf{R}}\textsub{QM}$, e.g., the centroid of the QM region. 
For each MM atom, it is possible to define a vector $\mathbf{R}_i = \mathbf{R}\textsub{MM,\,$i$} - \bar{\mathbf{R}}\textsub{QM}$
which is used in the calculation of the long-range interaction energy
\makeatletter
\if@twocolumn
    \begin{equation}
        \label{eq:lr}
        \begin{split}
        V\textsub{QM/MM}^\text{lr} &= \sum_{i=1}^{N\textsub{MM}\textsup{lr}} q_i 
        \left(
        \frac{q\textsub{QM}}{R_i} +\frac{\mathbf{R}_i^\mathrm{T}  \boldsymbol{\mu}\textsub{QM}}{R_i^3}
        \right.\\
        &\quad\left.+\frac{\mathbf{R}_i^\mathrm{T} \boldsymbol{\Theta}\textsub{QM} \mathbf{R}_i}{R_i^5}
        + \dots
        \right) \, ,   
        \end{split}
    \end{equation}
\else
    \begin{equation}
        \label{eq:lr}
        V\textsub{QM/MM}^\text{lr} = \sum_{i=1}^{N\textsub{MM}\textsup{sr}} q_i 
        \left(
        \frac{q\textsub{QM}}{R_i} +\frac{\mathbf{R}_i^\mathrm{T}  \boldsymbol{\mu}\textsub{QM}}{R_i^3} +\frac{\mathbf{R}_i^\mathrm{T}  \boldsymbol{\Theta}\textsub{QM}  \mathbf{R}_i}{R_i^5}
        + \dots
        \right) \, ,   
    \end{equation}
\fi
\makeatother
where $q\textsub{QM}$, $\boldsymbol{\mu}\textsub{QM}$, and $\boldsymbol{\Theta}\textsub{QM}$ are respectively the point charge, dipole vector, and quadrupole matrix of the QM region, given by the expressions
\begin{subequations}
\label{eq:expr_multipoles}
    \begin{equation}
    \label{eq:expr_q}
    q\textsub{QM} =  \int \mathrm{d} \mathbf{r}\,
    \rho(\mathbf{r})  + 
    \sum_{j=1}^{N\textsub{QM}}  Z_{j} \, ,    
    \end{equation}
    \begin{equation}
    \label{eq:expr_mu}
    \boldsymbol{\mu}\textsub{QM} =  \int \mathrm{d} \mathbf{r}\,   
    \rho(\mathbf{r}) \left(\mathbf{r} - \bar{\mathbf{R}}\textsub{QM}\right) + 
    \sum_{j=1}^{N\textsub{QM}}  Z_{j}  \left(\mathbf{R}\textsub{QM,\,$j$} - \bar{\mathbf{R}}\textsub{QM}\right) \, ,
    \end{equation}
    \begin{equation}
    \label{eq:expr_Q}
    \begin{split}
        \boldsymbol{\Theta}\textsub{QM} &=  \int \mathrm{d} \mathbf{r}\,   
    \rho(\mathbf{r}) \left[\frac{3}{2}\left(\mathbf{r} - \bar{\mathbf{R}}\textsub{QM}\right)\left(\mathbf{r} - \bar{\mathbf{R}}\textsub{QM}\right)^\mathrm{T}  - \frac{1}{2} \left|\mathbf{r} - \bar{\mathbf{R}}\textsub{QM}\right|^2\mathbf{I}\right]\\&+ 
    \sum_{j=1}^{N\textsub{QM}}  Z_{j} 
    \left[\frac{3}{2}\left(\mathbf{R}\textsub{QM,\,$j$} - \bar{\mathbf{R}}\textsub{QM}\right)\left(\mathbf{R}\textsub{QM,\,$j$} - \bar{\mathbf{R}}\textsub{QM}\right)^\mathrm{T}  - \frac{1}{2}\left|\mathbf{R}\textsub{QM,\,$j$} - \bar{\mathbf{R}}\textsub{QM}\right|^2\mathbf{I}\right] \, .
    \end{split}
    \end{equation}
\end{subequations}
The implementation of this scheme in MiMiC generalizes the original formulation\cite{laio2002hamiltonian} by allowing an open-ended multipole expansion in Eq.~\ref{eq:lr}.
This gives the users the possibility to select an expansion order that effectively compensates for a smaller short-range region that needs to be treated explicitly (Eq. \ref{eq:sr}).
By increasing the precision of the long-range contributions, it is thus possible to maintain the desired simulation accuracy but with a significant reduction of the overall computational cost.\cite{olsen2019mimic}

Along the dynamics, MiMiC sorts the particles of the system in sr and lr groups based on their distance from $\bar{\mathbf{R}}\textsub{QM}$ at user-defined intervals. 
This operation can be expensive, in particular for large systems, but it does not need to be performed at every MD step.
The optimal sorting rate depends on the system details and how fast the particles move.
Generally, the sorting is not a bottleneck in MiMiC, and overall using this sr/lr scheme allows for a net gain in performance.
Moreover, the sorting routine will be parallelized in a future MiMiC release.
For a detailed theoretical treatment of the electrostatic embedding scheme implemented in MiMiC, we refer to previous publications\cite{olsen2019mimic, antalik2024mimic}.

\subsection{OpenMM}
OpenMM\cite{eastman2013openmm, eastman2017openmm, eastman2023openmm} was initially developed as a library for MM calculations targeting GPUs.
Yet, over time, it has evolved into a sophisticated simulation package with many unique and powerful features, such as the support for polarizable force fields and implicit solvation models\cite{eastman2013openmm, eastman2017openmm} and, more recently, machine learning potentials.\cite{eastman2023openmm}
It has been designed to be simple, extensible, and easy to use while also offering excellent performance on different architectures, with a particular focus on GPUs.

OpenMM is based on a layered architecture consisting of four major components: the core OpenMM library, a set of \emph{platforms}, the \emph{application layer}, and various plugins.
The core library, written in C++, defines the types of calculations supported by OpenMM.
These include, e.g., bonded and nonbonded forces, with specific terms for the AMOEBA forcefield, as well as Generalized Born implicit solvent models.
The core library also includes a set of MD integrators, such as Verlet and Langevin (implemented with fixed and variable time step versions), Brownian and Nosé--Hoover integrators, and also a multiple time step Langevin integrator\cite{eastman2023openmm}.  
The variable time step integrators adjust the step size along the dynamics to keep the integration error below a user-specified tolerance.\cite{eastman2013openmm}
The actual calculations are implemented within the platforms, where the computations defined by the core OpenMM library are implemented to perform efficiently on specific hardware.
Current platforms include a \emph{Reference Platform}, serving as a reference when writing other platforms, a \emph {CPU Platform} which provides high performance when running on conventional CPUs, an \emph{OpenCL Platform} which uses the OpenCL framework and performs well on different types of GPUs and multicore CPUs, and a \emph{CUDA Platform}, specifically targeting NVIDIA GPUs.
The core OpenMM library and the platforms take care of the essential tasks for running MD simulations, i.e., computing forces and energies, and integrating the equations of motion. 
Other useful features for running MD simulations are contained in the application layer, which consists of a set of Python libraries that allow users to, e.g., load input structures, build FF topologies, or save trajectories and other data along the MD in standardized formats.
Finally, to ensure extensibility and modularity, plugins can be packaged as dynamically linked libraries, adding new features to OpenMM, including, for example, new types of interactions and integration algorithms.

\section{\label{sec:implementation}Implementation Details}
The OpenMM--MiMiC interface presented in this paper consists of a standalone program written in C++ that integrates the OpenMM C++ core library into the MiMiC loop, which is described in more detail below.
As this program follows a typical outline of a MiMiC interface, a virtually identical approach can be taken to write an interface with any external code.
In this work we used MiMiC release v0.2.0\cite{mimic:0.2.0} and MCL release v2.0.2\cite{mcl:2.0.2}, which support the use of CPMD v4.3\cite{cpmd_free} and GROMACS v2021.6\cite{gromacs:2021.6} as a QM and an MM client program, respectively.
To introduce OpenMM into the MiMiC ecosystem, we used the recent major release 8.x\cite{openmm:8.0.0} of the C++ core library.

\makeatletter
\begin{figure}[ht]
        \centering
    \includegraphics[width=.5\textwidth]{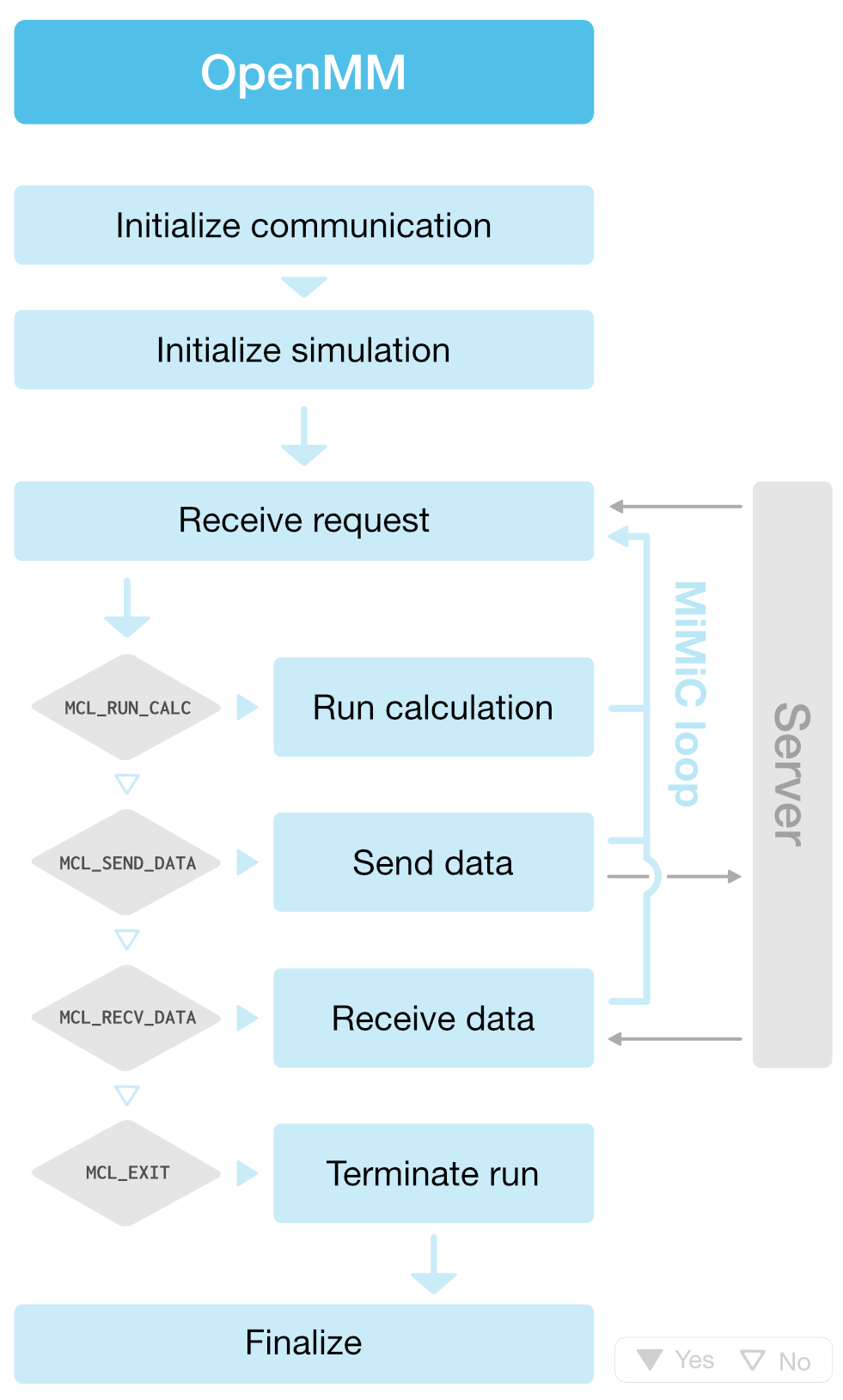}
    \caption{Schematic representation of the communication workflow in the OpenMM--MiMiC interface, where OpenMM acts as an external client program.
    Thick lines represent the workflow, while thin horizontal lines represent communication via MCL.}
    \label{fig:workflow}
\end{figure}
\makeatother

The workflow of the OpenMM interface program is illustrated in Fig.~\ref{fig:workflow}.
The interface program's main function starts with an initialization of MPI that is directly followed by an interception of the world communicator with an MCL function that returns a local 'world' communicator specific for OpenMM.
Then, it assumes the role of a client within the group of all concurrently running programs by invoking a handshake function from MCL, which concludes the initialization from the MiMiC side.

The next step is the OpenMM initialization, which requires setting up an OpenMM \verb|System| class that specifies generic properties of the system, such as the number of particles and the size of the simulation box.
This is done by reading a serialized System XML file containing force field parameters for the system, thus facilitating the system preparation for QM/MM with MiMiC.
This way, users can set up and equilibrate a system with classical MD and easily save the topology parameters as an XML file. 
The QM atoms in the system are identified by their indices provided by the user through an input file.
Their parameters in the FF topology are then adapted for a QM/MM simulation, as described below.
The coordinates of the system are read from a PDB file, which can be straightforwardly generated by OpenMM, e.g., at the end of the system equilibration at MM level.
Those coordinates are set as initial positions in an OpenMM \verb|Context| class, where state information for the simulation is stored, such as particle positions and velocities.

\begin{figure}
    \centering
    \includegraphics[width=0.5\textwidth]{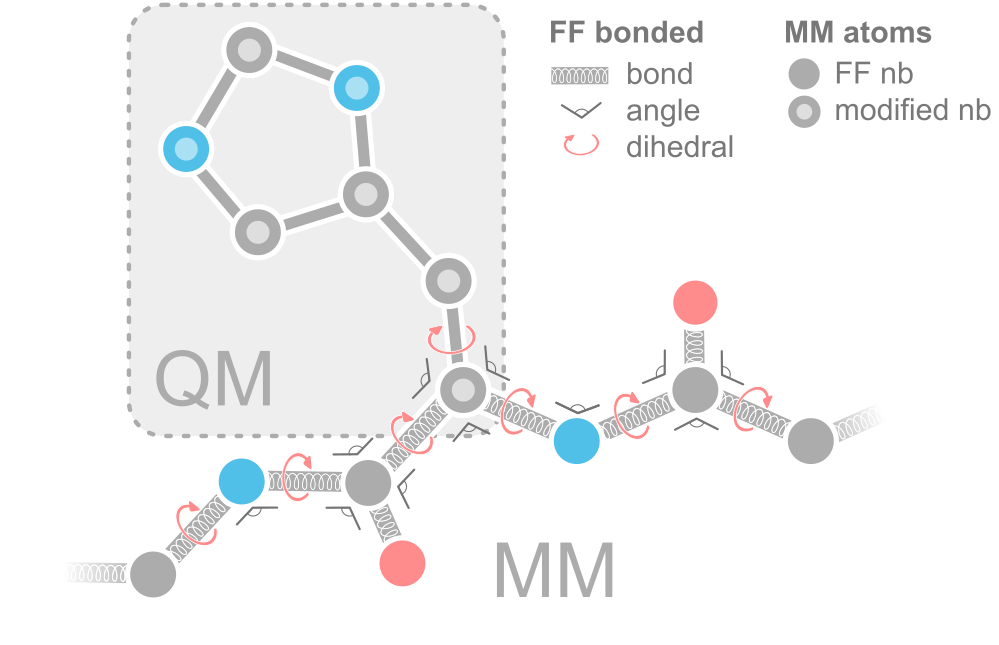}
    \caption{Schematic representation of the modifications introduced in the FF topology of the MM client program for the QM atoms in the system.
    Hydrogen atoms are not represented for clarity.
    Only bonded FF terms that are shown are included.
    Full circles indicate MM atoms whose topology is unchanged.
    Nonbonded (nb) FF parameters of the QM atoms are modified by zeroing their point charges and excluding vdW interactions between QM atoms but not between QM and MM atoms.}
    \label{fig:mm-topology-scheme}
\end{figure}

The FF parameters for the QM atoms are modified to avoid double-counting interactions (Fig.~\ref{fig:mm-topology-scheme}). 
Specifically, FF terms involving bond stretching, bending, and torsions solely between QM atoms are excluded. 
Similarly, point charges of QM atoms are zeroed and the vdW interactions between these atoms are excluded, while those involving vdW interactions between QM and MM atoms remain unchanged.

\begin{algorithm*}[ht]
\caption{The MiMiC loop in the program with examples of send and receive actions.}
\label{alg:example_algorithm}
\footnotesize	
\begin{lstlisting}[language=C++]
while (!isLastStep) {
  MCL_receive(&request, 1, TYPE_INTEGER, MCL_REQUEST, 0);
  if (request == MCL_SEND_ATOM_COORDS) {
    CurrentState = context.getState(OpenMM::State::Positions);
    const std::vector<OpenMM::Vec3>& Positions = 
                                       CurrentState.getPositions();
    std::vector<double> ConvertedCoordinates = 
                          convertCoordsToSend(Positions);
    MCL_send(ConvertedCoordinates.data(), ConvertedCoordinates.size(), 
             TYPE_DOUBLE, MCL_DATA, 0);
  }
  else if (request == MCL_RECV_ATOM_COORDS) {
    int numParticles = system->getNumParticles()
    std::vector<double> GetCoordinates(3*numParticles);
    MCL_receive(GetCoordinates.data(), 3*numParticles, 
                TYPE_DOUBLE, MCL_DATA, 0);
    std::vector<OpenMM::Vec3> ConvertedCoordinates = 
                                convertCoordsRecv(GetCoordinates);
    context.setPositions(ConvertedCoordinates); 
  }
  ...
  else if (request == MCL_EXIT) {
    isLastStep = true; 
  }
  else {
    Abort("Unrecognized MiMiC request!");
  }
}
\end{lstlisting}
\end{algorithm*}

Once the communication is initialized and the MM system is set up, the program enters the MiMiC loop, whose every iteration comprises two steps: it receives a request from the server and then executes the requested action.
This can be either an instruction to send or receive data, perform a calculation, or a request to terminate the simulation.
In case of send data requests, it is first necessary to fetch the data from OpenMM data structures, convert it to units compatible with MiMiC, and save it in an array.
The transmission of this data is then facilitated through a set of functions that constitute the MCL API. 
The same applies to receive data requests, though in a reversed order.
Two examples of such requests in the MiMiC loop, as implemented in our program, are presented in Alg.~\ref{alg:example_algorithm}.
When the program receives the \verb|MCL_SEND_ATOM_COORDS| request, the atom positions are fetched from the current OpenMM Context class, converted to the proper format, and sent via \verb|MCL_send|.
Similarly, when the \verb|MCL_RECV_ATOM_COORDS| request is received, the coordinates are received via \verb|MCL_receive|, converted, and set to the current Context.
Finally, the conditional loop is exited once the \verb|MCL_EXIT| request is received. 
After receiving this request, the program deallocates resources, terminates the MPI execution environment, and finally terminates.

\section{\label{sec:systems}Computational Details}
\makeatletter
\begin{figure}[ht]
        \centering
        \includegraphics[width=.5\textwidth]{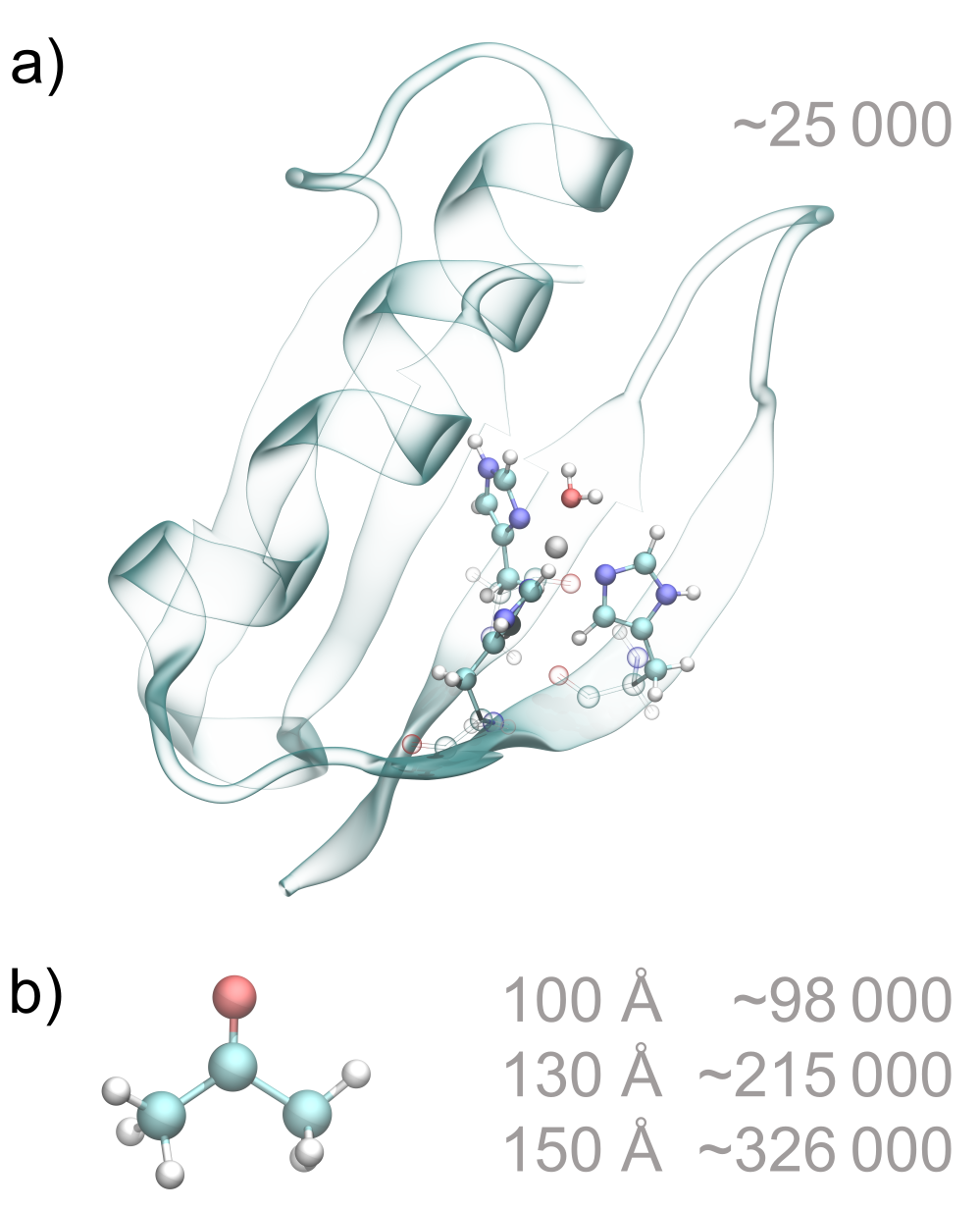}
        \caption{Systems simulated with QM/MM MD. 
        The QM subsystems are shown in ball-and-stick representation (full color) and the number of atoms in the system is reported in gray. 
        The water molecules (treated with MM) are omitted for clarity.
        a) GB1 protein mutant with protein backbone in cartoon representation, transparent. 
        b) Acetone molecule in water.
        For this system, three different box sizes have been used, with a different number of water molecules.}
        \label{fig:systems}
    \end{figure}
\makeatother

To validate and test the OpenMM--MiMiC interface program presented in this paper, we performed Born--Oppenheimer QM/MM MD simulations with several different systems.

First, we validated the new MM client introduced into the MiMiC ecosystem against its already available interface with GROMACS.
These simulations were performed on a protein system including a zinc-binding site, i.e., a GB1 metallo-mutant solvated in water \cite{bozkurt2018reprogramming,bozkurt2018genetic} (Fig.~\ref{fig:systems}a).
Its QM subsystem consists of the zinc ion, three coordinating histidine side chains, and one coordinated water molecule resulting in a total of \num{40} QM atoms.
The MM subsystem then consists of the rest of the protein, water molecules, and two \ce{Na+} counterions resulting in a total of \num{25936} atoms.
This is a good system for validation since, despite the small size, it presents a similar degree of complexity as typical biological systems and also contains three boundary atoms at the QM--MM interface (here described by monovalent pseudopotentials\cite{von2005variational}).

We also carried out a detailed performance comparison between OpenMM and GROMACS when used as MM client programs in a MiMiC-based QM/MM MD simulation and tested the impact of single- vs double-precision calculations.
Since we wanted to assess the performance of the MM part of the simulation, which in typical QM/MM applications is usually far from being the bottleneck, we opted for a system that can be easily increased in size.
Therefore, we used a simple solute--solvent system of a single acetone molecule consisting of \num{10} atoms treated at QM level solvated in MM water boxes of increasing size (Fig.~\ref{fig:systems}b).
In particular, we used cubic boxes of \qty{100}{\angstrom}, \qty{130}{\angstrom}, and \qty{150}{\angstrom} sides containing \num{98304}, \num{215235}, and \num{325980} atoms, respectively.
The rationale for selecting this set of systems was that they have a small, easy-to-solve QM subsystem embedded in MM subsystems of increasingly larger size. 
In this way, by assigning limited resources to the MM client program, the cost of calculating the MM contributions tends to become comparable with, or even higher than the one for the QM calculations.

For all QM/MM MD simulations, we used a time step of \qty{1}{\femto\second} and the BLYP functional\cite{Becke1988,Lee1988} with a plane-wave cutoff of \num{70}\,Ry to treat the QM atoms together with FF parameters for the MM subsystem from Ref.\citenum{olsen2019mimic}, i.e. OPLS/AA force field\cite{jorgensen2005potential} for the acetone molecule, AMBER ff14SB\cite{maier2015ff14sb} for the GB1 protein, and the TIP3P water model\cite{jorgensen1983comparison}.
The systems were initially equilibrated following a typical MM MD equilibration procedure, i.e., energy minimization, heating to \qty{300}{\kelvin}, short NVT equilibration, and longer NPT equilibration.
After switching to QM/MM, we annealed the systems to a few Kelvin for relaxation of the QM subsystem and reheated them again to \qty{300}{\kelvin}.
They were then maintained at the final temperature before switching to the NVE ensemble for a short MD run.
From these simulations, we extracted positions, velocities, and wavefunctions to start production runs, still in the NVE ensemble. 

To address the reproducibility of QM/MM MD simulations with GROMACS, we investigated the effect of using the option \texttt{reprod}.
This keyword, which is not activated by default, removes all sources of non-reproducibility and thus ensures results that are always reproducible for the same executable, hardware, shared libraries, input file, and parameters\cite{gromacs_manual_2021}.
However, all observables should converge to their equilibrium values even if this keyword is not used \cite{gromacs_manual_2021}.
In the simulations presented in the rest of the paper, we always explicitly specify when the keyword was activated.
Otherwise, the default setting was used, i.e., without the \texttt{reprod} option.

Part of the analysis has been performed using VMD\cite{HUMP96}, specifically the evaluation of the root-mean-square deviation (RMSD).

All simulations were performed on the Piz Daint cluster, hosted by the Swiss National Supercomputing Centre\cite{daint_cluster}.
It is a hybrid Cray XC40/XC50 system presenting Intel Xenon E5-2690 12-core CPUs and NVIDIA Tesla P100 GPU devices.
The hybrid CPU/GPU architecture allowed us to compare the impact of GPU-enabled MM client programs on QM/MM MD simulations with MiMiC.

\section{\label{sec:results}Results and Discussion}
In this section, we validate that OpenMM works as intended by performing QM/MM MD simulations with the new interface program and comparing it to the results obtained with GROMACS which has already been interfaced with the released version of MiMiC\cite{olsen2019mimic}.
In the rest of the paper, we present a detailed investigation of the possible performance gains from GPU support by the MM client program and the influence of single- vs double-precision calculations.

\subsection{Validation}
\begin{figure}[ht]
    \centering
    \includegraphics[width=0.48\textwidth]{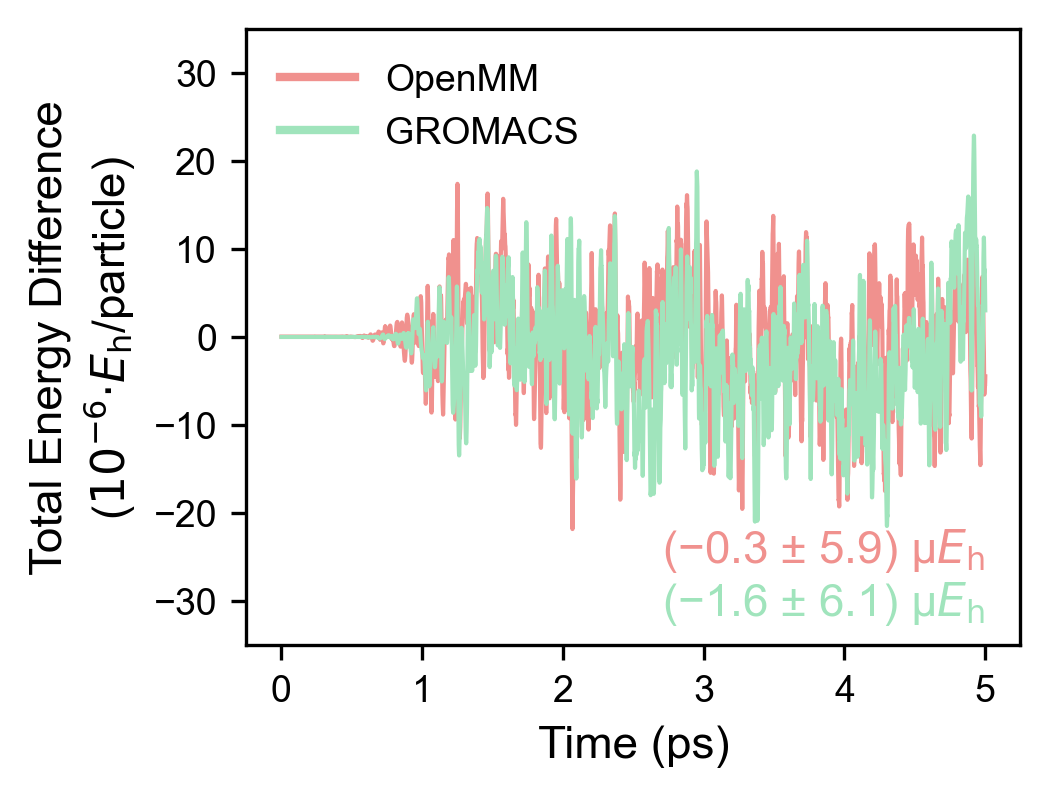}
    \caption{Differences in the total energy per particle between a reference GROMACS QM/MM MD simulation of the solvated GB1 protein mutant and a QM/MM MD simulation with the same initial positions and velocities, performed with OpenMM (red) and GROMACS (green) as MM client programs.
    The average values and standard deviations are also reported.}
    \label{fig:interface_validation_energy}
\end{figure}

\begin{figure}[ht]
    \centering
    \includegraphics[width=0.48\textwidth]{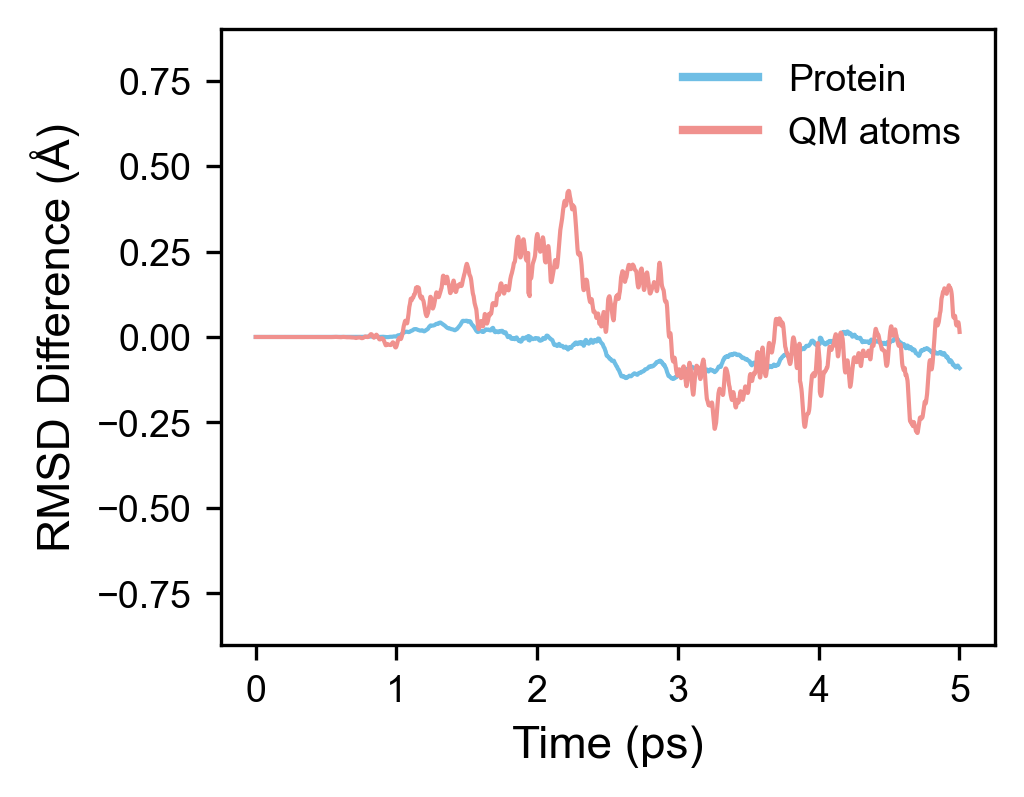}
    \caption{Difference in the RMSD between two QM/MM MD simulations of the solvated GB1 protein mutant with GROMACS and OpenMM as MM client programs, starting from the same initial positions and velocities. 
    The RMSD is computed for the atoms of the protein (green) and for the atoms of the QM subsystem (blue).}
    \label{fig:interface_validation_RMSD}
\end{figure}

Using the GB1 mutant system described above, we validated the new interface by comparing results using OpenMM as MM client program with the released version of MiMiC, which uses GROMACS. 
In both cases, we used CPMD as QM client program with the same input file, to assess the effect of using a different MM client program.
We performed two QM/MM MD runs of \qty{5}{\pico\second} for both clients, each time starting from the same initial positions and velocities, and we monitored different quantities along the dynamics.

First, we monitored ensemble quantities such as total energy and temperature.
With OpenMM, we were able to reproduce the results obtained from the simulations performed with GROMACS, with no difference in the monitored quantities for the first $\sim$\qty{0.7}{\pico\second} of the simulation (\num{7000} MD steps).
After that, these differences start to fluctuate around their average values which are \qty[separate-uncertainty=true]{-0.3\pm5.9}{\micro\hartree} for the total energy per particle (see Fig.~\ref{fig:interface_validation_energy}), and \qty[separate-uncertainty=true]{0.05\pm1.82}{\kelvin} for the temperature.
This suggests that the two quantities converged to the same equilibrium values. 

To show that the magnitude of the fluctuations is reasonable, we further calculated the differences between the reference GROMACS run, with the \texttt{reprod} keyword activated, and an independent simulation performed with GROMACS, without asserting reproducibility.
All the simulations performed with OpenMM reached the same ensemble averages and presented similar standard deviations as those performed with GROMACS, confirming the agreement between the interfaces for two different MM client programs (average values reported in Supporting Information).
In reference to reproducibility, OpenMM and GROMACS (when using the \texttt{reprod} keyword) generated the same results with the same version of the programs, with no difference in the total energy and the temperature for the entire simulation initiated from the same starting point.

Furthermore, to verify that not only ensemble quantities are consistent, we also analyzed the RMSD for the trajectories obtained with the two clients, with the results reported in Fig.~\ref{fig:interface_validation_RMSD}.
These show that the difference in the RMSDs between a MiMiC-based simulation with OpenMM and GROMACS for the atoms of the protein or the ones of the QM subsystem only is zero for $\sim$\qty{1}{\pico\second}.
After that, the differences in RMSD values begin to fluctuate, but with deviations remaining lower than $\sim$\qty{0.5}{\angstrom}.
Also in this case, we performed additional QM/MM MD simulations with GROMACS without the \texttt{reprod} keyword which showed similar fluctuations as the ones observed for ensemble quantities, with deviations lower than $\sim$\qty{0.5}{\angstrom} for the QM atoms.

\subsection{Single-node performance comparison}
\begin{figure*}[ht]
    \centering
    \includegraphics[width=\linewidth]{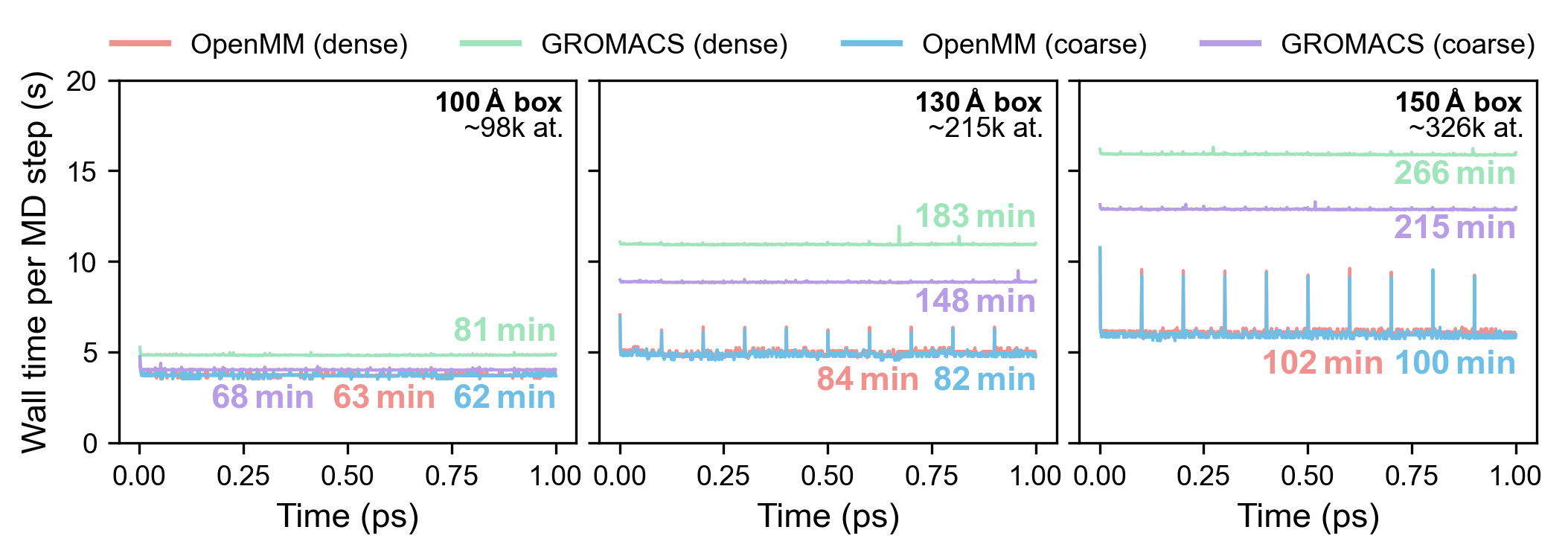}    
    \caption{Wall time per QM/MM MD step for the solvated acetone system in boxes of increasing size, with total run time printed in matching colors.
    The plotted values are the averages of three independent runs for each MM client program and PME setting (dense or coarse grid).
    Total wall time per simulation reported in the plot.}
    \label{fig:performance_comparison_averages}
\end{figure*}

The evaluation of classical FF energies and forces usually does not present a computational bottleneck in QM/MM MD simulations.
However, for very large systems with a small QM subsystem, the performance of the MM client program may affect the overall timing.
This becomes more relevant with the current development of more efficient QM programs, facilitated by better use of GPUs, or in simulations involving semi-empirical QM methods, which are notably much faster than ab initio methods.
To stress-test the two interfaced MM client programs, we performed simulations using a single node, where one MPI process was assigned to the MM client program, and the rest to MiMiC and CPMD for the more expensive QM/MM and QM calculations, both supporting only CPU calculations.
Efficient QM/MM simulations on a single node can be particularly attractive in specific contexts, e.g., for enhanced sampling techniques\cite{fedorov2013efficient} such as umbrella sampling, where multiple independent simulations of the same system need to be performed, or, even more, replica-exchange molecular dynamics, where those simulations are run in parallel and data are exchanged among them.
Performing QM/MM simulations efficiently on a single computer node can also be beneficial for development purposes or for users running on local workstations.

One of the benefits of using OpenMM is that, unlike GROMACS, it supports calculations in double precision on GPUs and not only on CPUs.
Hence, in these tests, OpenMM had not only access to a single CPU core but also to the GPU of the node.
Since QM programs almost exclusively use double precision, it is logical to use it also in the MM client program when running QM/MM MD simulations.
From the test system analyzed in this work, it is clear that the choice of numerical precision in the MM client program directly impacts the simulation reproducibility in the context of QM/MM MD, with only double precision calculations ensuring full reproducibility.

In this section, we employed systems that comprise a small QM subsystem, composed of a single acetone molecule, solvated in an increasing number of water molecules described by a classical FF.
To even out potential imbalances, we analyzed the time per MD step as an average for three independent runs for each program and grid used in the particle mesh Ewald (PME) method (Fig.~\ref{fig:performance_comparison_averages}).
Regarding the latter, we tested the Fourier grid parameters assigned by default in GROMACS (coarse), and OpenMM (dense) which was considerably finer.
The results presented in Fig.~\ref{fig:performance_comparison_averages} illustrate that with increasing MM system size, the performance of GROMACS, which ran on a single CPU core, starts significantly lagging behind the OpenMM performance.
Moreover, moving from a coarse to a dense grid for PME significantly increases the time per step in the case of GROMACS, while the performance of OpenMM is almost unaffected.

Interestingly, as the system size increases, regular peaks appear for the simulations using OpenMM.
They originate from sr/lr atom resorting in MiMiC, which is performed every 100 MD steps in these simulations.
Even though the time spent sorting atoms is the same in all runs, it does not affect the wall times in simulations with GROMACS as the sorting and MM calculation run concurrently, and the time for GROMACS to compute the MM contribution is slower than the sorting.
Also, with an increasing number of MM atoms and thus more time spent in sorting, the peaks become more prominent in the OpenMM runs.

Furthermore, the communication between OpenMM and MiMiC remains as efficient as for GROMACS.
A detailed breakdown of timings shows that, for the same system, the only significant difference between MiMiC QM/MM MD simulations using GROMACS or OpenMM is due to the computation of MM forces and energies performed by the MM client programs.

\subsection{Single-precision calculations for the MM client}
\begin{figure*}[ht]
    \centering
    \includegraphics[width=\linewidth]{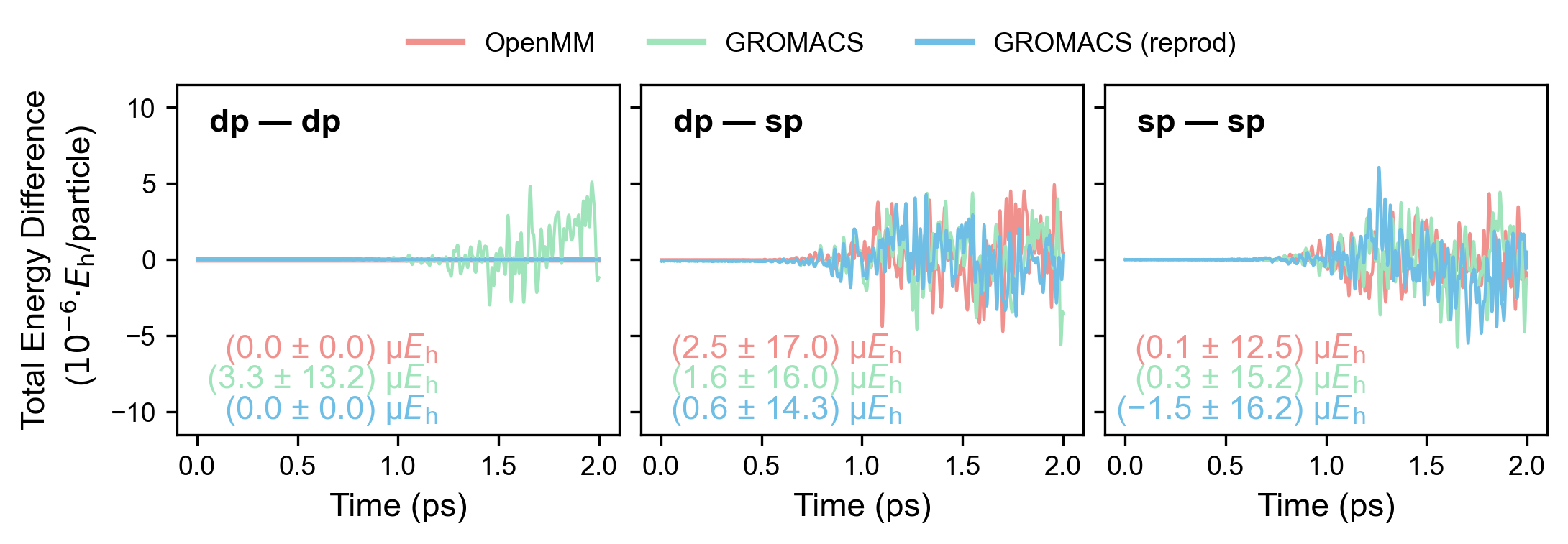}
    \caption{Differences in the total energy per particle between MiMiC-based QM/MM MD simulations of acetone in water using different MM client programs with double precision (dp) and single precision (sp). 
    The left plot shows the differences between two double-precision simulations (dp -- dp) performed with the same MM client program, the central plot between a double- and a single-precision simulation (dp -- sp), and the right plot between two single-precision simulations (sp -- sp). 
    The average values and standard deviations are also reported.}
    \label{fig:single_prec_comparison_averages}
\end{figure*}

In the performance comparison above, GROMACS had no access to a GPU because we restricted the simulations to double-precision floating-point arithmetics. 
However, it is possible to take advantage of GPUs in GROMACS as well, given that it is compiled with single-precision only.
Since MM programs are usually optimized for single precision, here we assess the impact of such a choice for the MM client program on the simulation accuracy, while the QM and QM/MM calculations are performed with double-precision accuracy.
For this, we used the most demanding system studied in the previous section, namely a single acetone molecule in a \qty{150}{\angstrom} water box with a dense PME grid, using a single node, where one CPU core was assigned to the MM client program, and the rest of the node to MiMiC and CPMD. 

Similarly, as for the validation above, the total energies and temperatures are compared in MD runs where the MM client program used either single- or double-precision floats to check if this change causes a significant loss in accuracy or reproducibility.
In addition, in the case of GROMACS, we again analyzed the effect of using the \texttt{reprod} keyword, which previously allowed us to obtain fully reproducible QM/MM MD trajectories.

The results from the simulation with OpenMM as the MM client program do not present any significant difference in terms of timings when moving from single- to double-precision calculations, thus resulting in the practically identical average total run time (\qty{101}{\minute} vs. \qty{102}{\minute}).
This is because even with double-precision, the MM calculation was already fast enough, thanks to the access to GPUs, with respect to the QM one, and did not represent a bottleneck for the overall QM/MM simulation.
On the other hand, in the simulation with GROMACS, we observed a significant gain in performance (owing to its access to GPUs), which is now on par with that of OpenMM (\qty{101}{\minute} or \qty{117}{\minute} with \texttt{reprod} enabled).

Concerning accuracy and reproducibility within the same version of the programs, switching to single-precision impacts the investigated quantities for both MM client programs (depending on the use of the \texttt{reprod} keyword, in the case of GROMACS).
The central and right-hand plots in Fig.~\ref{fig:single_prec_comparison_averages} show the differences in total energies between a single-precision simulation and another double- or single-precision simulation, respectively, starting with the same coordinates and velocities.
For comparison, the left plot shows the same difference between two independent double-precision simulations.
The differences exhibit similar behavior as in fully double-precision calculations with GROMACS with default settings, i.e. without the \texttt{reprod} keyword, with larger fluctuations starting after approximately \qty{0.75}{\pico\second}. 
Nevertheless, these fluctuations remain small and oscillate around zero during the entire simulation, suggesting that the ensemble averages from the different runs are still compatible.

In the case of OpenMM, for systems of similar size as the one tested in this section, there is no reason to use single precision, since double precision ensures fully reproducible results without becoming a bottleneck in the QM/MM calculation.
On the other hand, with GROMACS, full simulation reproducibility seems to be achievable only when using the \texttt{reprod} keyword in conjunction with double-precision, hence without the possibility of GPU offloading.
This can be useful in a development phase for code debugging, and also, e.g., for simulations of rare events.
In such cases, it is helpful to be able to reproduce those simulations for a more detailed analysis, e.g., by re-running the simulation while writing the output more frequently.

\section{Conclusions and Outlook}

We introduce OpenMM as an MM client program to the MiMiC ecosystem enabling mixed QM/MM multiscale simulations with classical energy and force contributions provided by OpenMM. 
We demonstrated that when using the same force field, QM/MM MD simulations with OpenMM closely reproduced the results of the corresponding runs with GROMACS, often with similar computational efficiency.
We also provided examples of potential situations in which the combination of MiMiC with OpenMM can present a significant advantage. 
One such case is when the QM subsystem is small and computationally less of a burden, combined with an MM subsystem for which FF components are slow to evaluate due to its large size or a limited amount of assigned resources.
Thanks to the GPU support for both double- and single-precision, OpenMM showed good performance in QM/MM simulations on a single node with GPU access. 
Similarly, GROMACS showed analogous performance only with single-precision, thanks to its GPU support for this type of calculation.
However, we also demonstrated that fully reproducible simulations can be obtained only when the MM client program uses double-precision and, in this case, OpenMM can provide faster time-to-solution, thanks to the access to GPU offloading for different numerical precisions.
Efficient QM/MM simulations on a single node are particularly attractive for cases where multiple parallel simulations need to be performed, such as in replica-exchange molecular dynamics, where QM/MM methods are still not widely applied due to their high computational cost.\cite{bondanza2020polarizable}

This new addition opens the possibility of choosing between different MM client programs for QM/MM simulations with MiMiC and, at the same time, gives the OpenMM community access to multiscale simulations in combination with current and future MiMiC-supported QM engines.
The modular structure of both frameworks opens numerous ways for new features to be introduced in MiMiC and vice versa.

Possible extensions include the support for polarizable force fields, available in OpenMM via the AMOEBA\cite{Ponder2010-nr} and Drude oscillator\cite{lamoureux2003modeling} models, to perform QM/MM MD simulations with a fully-polarizable approach that is currently being implemented in MiMiC\cite{antalik2024mimic, sonata_polqmmm}, as well as different ML potentials\cite{eastman2023openmm}, making it possible to implement, e.g., MM/ML embedding models.
Another possible direction towards the large-system simulations might be the use of coarse-grained (CG) force fields, such as Martini\cite{Marrink2013, Marrink2022}, to implement, e.g., CG/MM or CG/MM/QM models. 
Moving even further, OpenMM also supports different implicit solvent models, which can be integrated into different continuum model (CM) multiscale schemes, such as CM/QM or CM/MM/QM.

MiMiC is an innovative framework for multiscale simulations and, thanks to the addition of software packages like OpenMM, which share similar design philosophies in terms of flexibility, efficiency, and extensibility, it becomes even more general, opening up new paths for future research avenues.

\section*{Associated Content}
The source code of the MiMiC framework is free and open-source and is hosted on GitLab\cite{mimic-projects}. 
The code of the interface program will soon be available on the same GitLab page after code review. 
The preliminary version used within this study, together with all the data generated, is available on Zenodo under \url{https://zenodo.org/records/14796427}, including a Jupyter Notebook to reproduce the analysis and generate the plots and tables presented.

Supporting Information contains details on the systems and simulations, a detailed timing breakdown for QM/MM MD simulations for performance comparison, as well as results from each independent MD simulation used to produce the average results presented in the main text.

\section*{Author Information}
\subsection*{Notes}
The authors declare no competing financial interest.

\section*{Acknowledgments}
We gratefully acknowledge Sonata Kvedaravičiūtė and David Carrasco-Busturia for fruitful discussions during the implementation and testing phases of this project.
This work has been supported by the Swiss National Science Foundation (grant 200020-185092 and 200020-219440), and used computing time from the Swiss National Computing Centre CSCS.
JMHO gratefully acknowledges financial support from VILLUM FONDEN (Grant No. VIL29478).

{\footnotesize
\bibliography{bib}
}




\end{document}